# RetiFluidNet: A Self-Adaptive and Multi-Attention Deep Convolutional Network for Retinal OCT Fluid Segmentation

*Reza Rasti, Member, IEEE; Armin Biglari; Mohammad Rezapourian; Ziyun Yang; and Sina Farsiu, Fellow, IEEE*

*Abstract*— Optical coherence tomography (OCT) helps ophthalmologists assess macular edema, accumulation of fluids, and lesions at microscopic resolution. Quantification of retinal fluids is necessary for OCT-guided treatment management, which relies on a precise image segmentation step. As manual analysis of retinal fluids is a time-consuming, subjective, and error-prone task, there is increasing demand for fast and robust automatic solutions. In this study, a new convolutional neural architecture named RetiFluidNet is proposed for multi-class retinal fluid segmentation. The model benefits from hierarchical representation learning of textural, contextual, and edge features using a new self-adaptive dual-attention (SDA) module, multiple self-adaptive attention-based skip connections (SASC), and a novel multi-scale deep self-supervision learning (DSL) scheme. The attention mechanism in the proposed SDA module enables the model to automatically extract deformation-aware representations at different levels, and the introduced SASC paths further consider spatial-channel interdependencies for concatenation of counterpart encoder and decoder units, which improve representational capability. RetiFluidNet is also optimized using a joint loss function comprising a weighted version of dice overlap and edge-preserved connectivity-based losses, where several hierarchical stages of multi-scale local losses are integrated into the optimization process.

The model is validated based on three publicly available datasets: RETOUCH, OPTIMA, and DUKE, with comparisons against several baselines. Experimental results on the datasets prove the effectiveness of the proposed model in retinal OCT fluid segmentation and reveal that the suggested method is more effective than existing state-of-the-art fluid segmentation algorithms in adapting to retinal OCT scans recorded by various image scanning instruments.

*Index Terms*— Medical Image Segmentation, Convolutional Neural Network, Retinal Disease, Fluid Segmentation.

## I. INTRODUCTION

THE retina in human eyes receives focused light and transforms it into neural signals. Located in the central region of the retina, the macula is responsible for detecting color, light intensity, and visual details. Optical coherence tomography (OCT) imaging captures 3-D images at micrometer resolution and provides cross-sectional scans for the retinal structure and its biological tissues [1]. In clinical ophthalmology, the OCT imaging technique has become the preferred scanning method for the macula because of its high resolution and visualization ability for the anatomical constituent layers of the photoreceptors [2], [3].

Diabetic macular edema (DME) is a major cause of blindness and visual impairments in patients with diabetic retinopathy (DR). Quality degradation of vision due to DME is related to damaged blood vessels and fluid accumulation between retinal layers. Thus, to properly manage the treatment of DME and prevent its progression to blindness, DME patients' retinal layer morphology and fluid accumulation must be thoroughly monitored [4], [5]. For the accurate assessment of the macula and fluid regions, precise segmentation of retinal OCT images is required [6]. Manual segmentation, although sufficient, is a time-consuming, subjective, and error-prone task; hence, fast and robust automatic alternatives are desired.

To analyze macular fluids automatically and accurately, different image processing and machine learning algorithms have been developed [7]–[9]. Of the advanced solutions in the field, deep learning provides the advantage of learning discriminative representations to obtain acceptable segmentation performance [10]–[12]. A review of the related works is presented in Section II.

Despite these recent advances, accurate retinal fluid segmentation in macular OCT imaging remains a challenging task due to variations in the structure of the retina, disease-related variations in fluid shape and size, blurred boundaries of the suspected regions, different levels of disruptive speckle noise, and low contrast of the OCT image data. These difficulties give rise to imperfect segmentation, especially for near-edge regions, and low spatial coherence. Moreover, the lack of sufficient fluid patterns with pixel-level annotation complicates the fluid segmentation task. In the present study, to address these challenges and improve performance, we propose a novel self-adaptive and multi-level attention convolutional U-

Manuscript received Nov X, 2021; revised X, 2021; accepted X, 2021. Date of publication X, 2021; date of current version X, 2021. This work was supported by the Department of Biomedical Engineering, University of Isfahan. (Corresponding author: Reza Rasti)

R. Rasti, A. Biglari, M. Rezapourian are with the Department of Biomedical Engineering, Faculty of Engineering, University of Isfahan, Isfahan, Iran (e-mail: r.rasti@eng.ui.ac.ir; armin.biglari@mehr.ui.ac.ir; mrezapourian@mehr.ui.ac.ir). R. Rasti is also a research collaborator with the VIP Lab, Department of Biomedical Engineering, Duke University, Durham, NC 27708 USA (e-mail: reza.rasti@duke.edu).

Z. Yang and S. Farsiu are with the Department of Biomedical Engineering, Duke University, Durham, NC 27708 USA. S. Farsiu is also with the Department of Ophthalmology, Duke University Medical Center, Durham, NC 27708 USA (e-mail: ziyun.yang@duke.edu; sina.farsiu@duke.edu).

...

shaped network called RetiFluidNet. Our main technological contributions in this work are summarized as follows:

(1) A novel self-adaptive, dual-attention (SDA) module is proposed and designed to jointly capture the relevant contextual information in the spatial and channel domains. The SDA module imposes pixel-group attention on input tensors by introducing an attention mechanism that accounts for spatial-channel interdependencies efficiently and adaptively. With only two trainable free parameters, the module can weigh and highlight spatial and channel information adaptively in an automatic and end-to-end manner to improve the attention process performance.

(2) A novel self-adaptive, attention-based skip connection (SASC) is proposed and inserted between the encoder and decoder counterpart blocks using the SDA module. The SASC paths improve the ability of the overall model to learn global and multi-semantic contextual representative features.

(3) By combining SDA and SASC modules, a novel convolutional architecture, RetiFluidNet, is designed based on the U-Net structure and used for fluid segmentation in macular OCT images.

(4) We further proposed and evaluated a deep self-supervision learning (DSL) scheme for the model's optimization in our multi-class segmentation problem. A new joint loss function comprising a multi-scale weighted version of dice overlap losses and edge-preserved connectivity-based losses is introduced. Several hierarchical stages of local losses are computed according to the consecutive decoder blocks' outputs and merged into the joint loss function. For this, multi-scale local losses of dice loss components (DLC) and connectivity-based loss components (CLC) are used for texture- and edge-specific feature enhancement, respectively. More specifically, in CLC, local connectivity masks are used together with annotated masks as labels for effective modeling of inter-pixel relationships and edge feature enhancement for each class of fluids. Finally, a linear combination of DLC and CLC terms is integrated into the optimization process to compute the overall objective function and obtain enhanced pixel-wise segmentation performance.

(5) We also conducted extensive experiments and ablation studies using three publicly available retinal fluid segmentation datasets (RETOUCH [13], OPTIMA [14], and DUKE [15]) to evaluate the effectiveness of the proposed model. Compared with other baselines and state-of-the-art deep learning-based fluid segmentation methods, RetiFluidNet performance was significantly better with comparative and acceptable execution time.

The remainder of this paper is organized as follows: Section II describes the latest research studies on the segmentation of fluid regions in retinal OCT images. Section III introduces the materials and formulates our proposed method in detail. Section IV shows the experimental study and results. Section V discusses our findings in this study and draws potential future research directions. Finally, in section VI, the study's conclusion is presented.

## II. RELATED WORK

Segmentation of disease-affected regions in retinal OCT imaging is an active and demanded area of research. Many machine learning methods have been proposed to segment retinal fluids and layers [2], [16], [17], and classify retinal diseases [18]–[21]. Recently, several deep-learning techniques have been applied to segment retinal lesions and layers. U-Net [22], FCN [23], Seg-Net [24], Deeplabv3+ [25], and U-Net++ [26] are among the most common and interesting deep learning architectures developed for image segmentation and used in medical applications. Table 1 summarizes recent studies that have applied deep learning models to retinal fluid segmentation.

Fluid segmentation algorithms are often affected by a high rate of false detections in the background regions [27], [28]. To overcome this issue, Chen et al. [29] and Fu et al. [30] have suggested the use of larger convolutional kernel sizes in convolutional neural network (CNN)-based methods to benefit from larger receptive fields, while [31], [32] studies have proposed using weighted versions of loss functions in the optimization process. Since enlarged receptive fields and weighted loss functions cannot be dynamically adjusted to fit fluid targets of different sizes, a variety of attention mechanisms have been proposed to address this challenge and develop scale-aware models [27], [33], [34].

In the study of human vision, the attention mechanism is often referred to as the process by which humans selectively focus on some parts of visual information while ignoring others, which results in enhanced visual information processing. To improve the performance of deep learning-based segmentation models, several attention mechanisms have been suggested and used in recent years [28], [34]. W. Liu et al. [27] proposed an attention-based method for fluid segmentation, and they integrated multi-scale inputs, multi-scale side-outputs, and attention methods into U-Net++ for better performance. In a study by X. Liu et al. [34] an attention-based U-Net model was developed for intraretinal cystoid fluid (IRC) segmentation in DME images with an attention structure that automatically considers suspected regions. Feng et al. [35] proposed the CPFNet model which explored global and multi-scale contextual information mainly through a soft spatial attention mechanism that learned features across different receptive fields using dilated convolutions with shared weights. They also tested the model's performance for automated segmentation of retinal edema lesions on the AI-challenger 2018 dataset with acceptable performance. Recently, Wang et al. [36] developed the MsTGANet model, a multi-scale transformer global attention network for drusen segmentation in retinal OCT images that combines non-local and multi-semantic information.

In this study, we present a multi-attention and self-adaptive neural structure that adaptively tunes the attention processes based on an end-to-end training scheme to reduce the number of false detections. Our model takes into account multi-scale textural, contextual, and edge information for better segmentation of noisy and blurry retinal fluid regions in OCT images. The suggested CNN architecture can efficiently manage large variations in location, size, and shape of diverse pathological fluid lesions. The efficiency of our proposed model is supported by extensive empirical results in Section IV.

TABLE 1. SUMMARY OF THE RECENT DEEP LEARNING METHODS APPLIED TO RETINAL FLUID SEGMENTATION

| Reference | Data | | | Pre-Processing | | | | Model | Validation Method | Evaluation Metric | Results* (%) |
|---|---|---|---|---|---|---|---|---|---|---|---|
| | Dataset: Scanner | Sample Size | Fluid Type | De-noise | Data Aug | Layer Seg | Crop-Resize | | | | |
| Roy et al. [2] | DUKE: Spectralis | 10 | IRC | | ✓ | ✓ | ✓ | ReLayNet, 2D | 8-F CV | DSC | 81 |
| Alsaih et al. [3] | RETOUCH, OPTIMA, DUKE: Cirrus, Nidek, Spectralis, Topcon | 112 +30 +10 | All | ✓ | ✓ | | ✓ | Deeplabv3+ext. patch, 2.5D | 3-F CV H.O H.O | DSC | RETOUCH:80,-,82, 86; OPTIMA:72,62, 70,78; DUKE:-,-,80,- |
| Alsaih et al. [5] | RETOUCH, OPTIMA: Cirrus, Nidek, Spectralis, Topcon | 112 +30 | All | ✓ | ✓ | | ✓ | Pa.Deeplabv3+, 2D | 3-F CV | DSC | RETOUCH:84,-,79,82; OPTIMA:-,63,-,- |
| Chiu et al. [15] | DUKE: Spectralis | 10 | IRC | ✓ | ✓ | ✓ | ✓ | Kernel Reg., 2D | 10-F CV | DSC | 53 |
| W.Liu et al. [27] | RETOUCH: Cirrus, Spectralis, Topcon | 112 | IRF, SRF, PED | ✓ | | | ✓ | MDAN U-Net, 2D | H.O | DSC | 67, 69, 65 |
| X.Liu et al. [34] | OPTIMA: Cirrus, Nidek, Spectralis, Topcon | 30 | IRC | | ✓ | | ✓ | Att-U-Net, 2D | 4-F CV | DSC | 83, 80, 84, 84 |
| Feng et al. [35] | AI Challenger 2018 | 83 | REA, SRF, PED | | ✓ | | ✓ | CPFNet, 2.5D | 5-F CV | DSC | 80 |
| Lee et al. [37] | Local Data: Spectralis | 30 | IRF | | ✓ | | | U-Net, 2D | H.O | DSC | 75 |
| Lu et al. [38] | RETOUCH: Cirrus, Spectralis, Topcon | 112 | IRF, SRF, PED | | ✓ | ✓ | | U-Net, 2D | L.O.O CV | DSC | 71, 77, 64 |
| Venhuizen et al. [39] | OPTIMA: Cirrus, Nidek, Spectralis, Topcon | 30 | IRC | | | | ✓ | CNN, 2D | H.O | DSC | 65, 66, 83, 81 |
| Girish et al. [40] | OPTIMA: Cirrus, Nidek, Spectralis, Topcon | 30 | IRC | ✓ | ✓ | | ✓ | U-Net, 2D | H.O | DSC | 63, 75, 74, 77 |
| Gopinath et al. [41] | OPTIMA: Cirrus, Nidek, Spectralis, Topcon DUKE: Spectralis | 30 | IRC | ✓ | ✓ | ✓ | ✓ | U-Net, 3D | H.O | DSC | OPTIMA:71, 73, 64, 76; DUKE:69 |
| Reis et al. [42] | RETOUCH: Cirrus, Spectralis, Topcon | 112 | IRF, SRF, PED | | ✓ | | ✓ | Semi-Weakly Supervised U-Net, 2D | 10-F CV | IoU | 51, -, 42 |
| Ren et al. [43] | RETOUCH: Cirrus, Spectralis, Topcon | 112 | IRF, SRF, PED | ✓ | | | ✓ | CGAN, 2D | H.O | DSC; IoU | 67, 62, 52; 51, 48, 43 |
| Mahapatra et al. [44] | RETOUCH (Pathological Images): Cirrus, Spectralis, Topcon | 112 | IRF, SRF, PED | | ✓ | ✓ | | GeoGAN, 2D | H.O | DSC; HD | Average=90; 7.9 |
| Schlegl et al. [45] | Local Data: Cirrus, Spectralis | 1200 | IRC, SRF | | | ✓ | ✓ | U-Net, 2D | Cirrus: 4-F CV Spect: 10-F CV | Pr; Re | 80, 86; 83, 75 |
| Yadav et al. [46] | Local Data: Cirrus, Spectralis, Topcon | 112 | IRF, SRF, PED | ✓ | | | ✓ | U-Net, 2D | 8-F CV | DSC; AUC | 72, 68, 66; 84, 82, 87 |
| Hassan et al. [47] | RETOUCH: Cirrus, Spectralis, Topcon | 112 | IRF, SRF, PED | ✓ | | ✓ | ✓ | RFS-Net, 2D | H.O | DSC; IoU | 75, 79, 80; 61, 65, 67 |

IRF=Intraretinal Fluid, SRF=Subretinal Fluid, PED=Pigment Epithelial Detachment, IRC=Intraretinal Cystoid Fluid, REA: Retinal Edema Area, DSC=Dice Coefficient Score, IoU=Intersection over Union, CV=Cross-Validation, H.O=Hold Out, L.O.O=Leave-One-Out, Pr=Precision, Re=Recall, HD=Hausdorff Distance, GAN= Generative Adversarial Network, Att=Attention. *The results are reported based on OCT scanners analyses, respectively. In [44], the results are only available for whole scanners.

## III. MATERIALS AND METHODS

In this section, we first describe the datasets used in this study for model evaluation and generalization, and then we present our data analysis pipeline and introduce in detail the pre-processing stage, model structure, and core attention components. Finally, we formulate the joint loss function and learning scheme integrated for optimizing the proposed model.

### A. Datasets

We used three publicly available datasets: the RETOUCH challenge dataset [13], OPTIMA challenge dataset [14], and DUKE dataset [15]. The RETOUCH dataset was collected to distinguish intraretinal fluid (IRF), subretinal fluid (SRF), and pigment epithelial detachment (PED) regions. This dataset contains three OCT scanners: Zeiss Cirrus, Heidelberg Spectralis, and Topcon. The OPTIMA dataset includes IRC regions for segmentation tasks. In addition to the previous three scanners, this dataset also has a Nidek scanner. The DUKE dataset was collected to segment DME-relevant fluids and only used the Spectralis scanner. Fig.1 provides sample illustrations of OCT B-scans for the RETOUCH, DUKE, and OPTIMA datasets. As the ground truth (GT) for the test set is not made available by the challenge organizers, the RETOUCH test data was not used in our study. Therefore, we considered the training set with cross-validation for evaluating our method and the baselines. Moreover, the models were further externally evaluated over the OPTIMA and the DUKE datasets. The training and testing sub-sets of the OPTIMA and DUKE datasets were both supplied with labels. The details of the datasets are summarized in Table 2.

In the present study, the RETOUCH challenge dataset was used to train and validate the proposed RetiFluidNet and baseline models. The OPTIMA and DUKE datasets were also used for comparison with the recent methods developed for the retinal fluid segmentation problem.

TABLE 2. DATASETS DESCRIPTION

| Scanner | Volume size | RETOUCH | | OPTIMA | | Volume size | DUKE |
|---|---|---|---|---|---|---|---|
| | | Tr | Te | Tr | Te | | |
| Cirrus | 512×1024×128 | 24 | 14 | 4 | 4 | - | - |
| Spectralis | 512×496×49 | 24 | 14 | 4 | 4 | 496×768×61 | 10 |
| Nidek | 512×512×128 | - | - | 3 | 3 | - | - |
| Topcon | 512×885×128 | 22 | 14 | 4 | 4 | - | - |
| GT Availability | - | - | Yes | No | Yes | Yes | - | Yes* |

Tr=train, Te=test, GT=Ground Truth, * For the DUKE dataset, only 11 mask labels per volume are available.

### B. Pre-Processing

All B-scan images in various volumes were initially resized to 256×256 pixels to obtain a unique field of view. Furthermore, a min-max normalization step was performed to map the intensity values of each B-scan between 0 and 1.

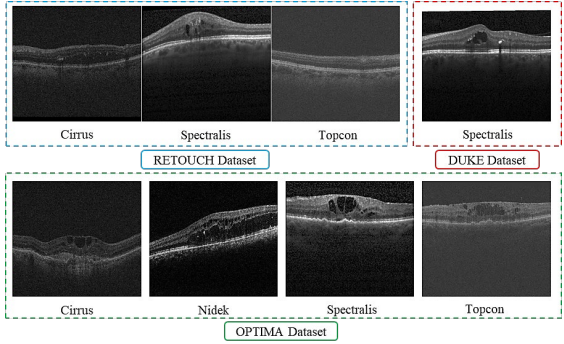

Fig. 1. B-scan samples of datasets.

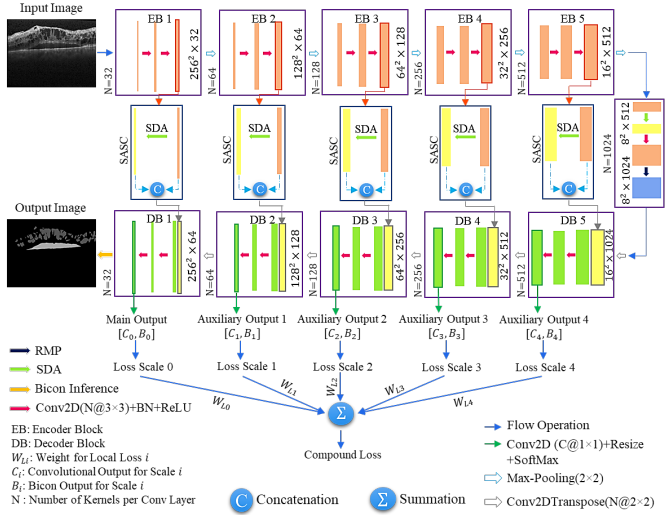

Fig. 2. The architecture of the proposed RetiFluidNet model. (Best viewed in color)

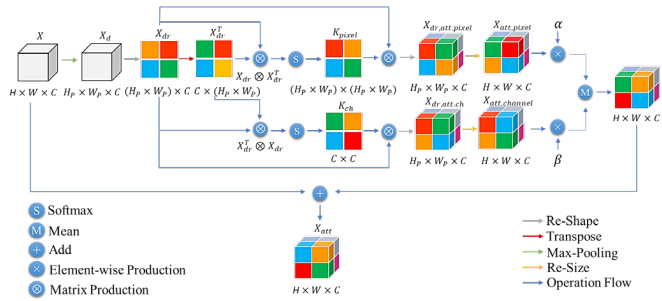

Fig. 3. Detailed structure of self-adaptive dual attention (SDA) block. (Best viewed in color)

### C. RetiFluidNet

Fig.2 demonstrates the proposed RetiFluidNet architecture, which adopts an encoder-decoder U-shaped topology as the basic framework. RetiFluidNet consists of five major components, including (1) encoder blocks, (2) SDA module, (3) SASC path, (4) residual multi-kernel pooling (RMP) module, and (5) decoder blocks.

*1) Encoder Blocks:* As shown in Fig.2 and like the U-Net structure, the encoder part of the proposed RetiFluidNet contains five blocks. Each block consists of two consecutive blocks of Conv2D-BN-ReLU layers with 3×3 kernels followed by a 2×2 Max-pooling operation. The Conv2D-BN-ReLU layers are used to extract the representative features in different stages, and the Max-pooling operation is adopted to abstract data and reduce the complexity of the model.

*2) SDA Module:* In the present work, we introduce a soft attention module inspired by the attention mechanisms in [48] to guide the network where to look for the fluid regions in the retinal layers. The proposed self-adaptive and dual attention module, SDA, enables the neural model to automatically extract deformation-aware representations at multiple levels. Several preceding studies have developed self-attention modules that demonstrate the positive impact of spatial- and/or channel attentions in improving various image analysis tasks [48]–[51]. Different from the earlier works, we extend the self-attention mechanism to the task of fluid segmentation, where the two types of attention are embedded in an integrated module to capture rich contextual relationships with intra-class compactness. The module aggregates spatial- and channel attention processes selectively and adaptively with only two trainable parameters for enhanced feature representations. The efficiency of the proposed SDA module is supported by experimental results in Section IV. The SDA module performs two attention mechanisms on the given tensor: a channel-wise attention process and a pixel-wise attention process. Fig.3 demonstrates the structure of the attention block. Using perturbed and augmented spatial attention, the proposed SDA module incorporates spatial and channel interactions for more efficient feature representation. The perturbation mechanism is embedded into the module where $K_{pixel}$ is computed in the processing pipeline. It extracts and attends spatial features from relevant parts of input feature maps (FMs). The representation process is then directed to generate relevant discriminative FMs by the perturbed region amplification. Meanwhile, the module calculates the $K_{ch}$ to attend various channel dimensions of the input FMs and efficiently encodes FM inter-channel interactions. By adopting higher values of weights over the signal flows, the model adaptively integrates the mechanisms to further converge on the relevant regions of the input tensor for better attention performance. By modeling rich contextual dependencies over local features, it significantly improves segmentation performance. The RetiFluidNet takes advantage of several SDA processes at different levels, gaining coarse and smooth attentive features from two intrinsic attention processes embedded into the SDA module design. The SDA module is used between the encoder and the decoder counterparts through SASC paths and is further embedded into the top of the final encoder block to enhance latent-space representation ability.

Let the input tensor for the SDA module be $X \in \mathbb{R}^{H \times W \times C}$, where $W$, $H$, and $C$ stand for width, height, and channel dimensions for the given tensor. The SDA generates two highlighted tensor maps: $X_{att,pixel} \in \mathbb{R}^{H \times W \times C}$, and $X_{att,channel} \in \mathbb{R}^{H \times W \times C}$, for pixel-wise attention and channel-wise attention, respectively. They are added to the input tensor using the following expression:

$$X_{att} = X + \frac{1}{2}(\alpha X_{att,pixel} + \beta X_{att,channel}) \quad (1)$$

Here, $X_{att} \in \mathbb{R}^{H \times W \times C}$ is the output tensor with multi-dimension attention, which represents the relevant contextual information in the spatial and channel domains. In practice, the SDA adds a highlighted version of the input, which is a weighted average of $X_{att,pixel}$ and $X_{att,channel}$. In this equation, $\alpha, \beta \geq 0$ are two trainable parameters that adaptively weigh the contribution of the pixel-wise and channel-wise attention mechanisms for driving the models' focus on more relevant representative parts of the input data. To generate $X_{att,pixel}$ and $X_{att,channel}$, the input tensor is fed to a Max-pooling layer by a pooling factor of $p$ and the resultant down-sampled tensor, i.e., $X_d \in \mathbb{R}^{H_p \times W_p \times C}$, is followed by a reshape layer that transforms the 3-D tensor into a 2-D matrix, $X_{dr} \in \mathbb{R}^{(H_p \times W_p) \times C}$, as shown here:

$$X_d = Maxpooling(X, p), \quad (2)$$
$$X_{dr} = Reshape(X_d), \quad (3)$$

The pixel-wise attention coefficient of $K_{pixel} \in \mathbb{R}^{(H_p \times W_p) \times (H_p \times W_p)}$ is calculated by:

$$K_{pixel} = Softmax\left(\frac{1}{\sqrt{H_p \times W_p}} X_{dr} \otimes X_{dr}^T\right), \quad (4)$$

In this equation, $T$ denotes transpose operation, the operator of $\otimes$ indicates matrix-multiplication, the normalization factor of $\frac{1}{\sqrt{H_p \times W_p}}$ limits the magnitude of the product output, and $Softmax(x_j) = \frac{\exp(x_j)}{\sum_{i=1}^{n} \exp(x_i)}$; $j = 1, 2, \ldots, n$ where $n$ is the number of matrix elements. The normalization factor, here, is used so that the Softmax function's input does not push it into regions where it has extremely small gradients in the optimization process. The matrix multiplication in Eq.4, amplifies the spatial similarity of $X_{dr}$ and $X_{dr}^T$ whereas the Softmax function yields context-aware attention coefficients by assigning greater probabilities to higher entropy regions.

The generated pixel-attention coefficient $K_{pixel}$ acts as a weighing function on $X_{dr}$ using Eq.5 and yields:

$$X_{dr,att.pixel} = K_{pixel} \otimes X_{dr}; \; X_{dr,att.pixel} \in \mathbb{R}^{(H_p \times W_p) \times C} \quad (5)$$

Subsequently, the resultant weighted matrix is reshaped back and resized to obtain $X_{att,pixel} \in \mathbb{R}^{H \times W \times C}$ feature tensor, given by:

$$X_{att,pixel} = Unpooling(Reshape(X_{dr,att.pixel}), p), \quad (6)$$

The second path in the SDA module (see Fig.3) exploits interdependencies between high-level FMs or channels of the input tensor. By explicitly modeling the correlations between high-level FMs, it emphasizes interdependent channels, leading to improved feature representation ability and better attention. Like Eq.4, channel-attention coefficients $K_{ch}$ can be mathematically written as:

$$K_{ch} = Softmax\left(\frac{1}{C} X_{dr}^T \otimes X_{dr}\right); \quad K_{ch} \in \mathbb{R}^{C \times C} \quad (7)$$

Before applying the Softmax function over the resultant matrix production, the scores are normalized using the number of FMs, i.e., $C$ value. Please note the order of multiplicand and multiplier in the matrix multiplication here. Then, $X_{dr,att.ch}$ is calculated by performing multiplication between $K_{ch}$ and $X_{dr}$ matrices using Eq.8.

$$X_{dr,att.ch} = K_{ch} \otimes X_{dr}; \; X_{dr,att.ch} \in \mathbb{R}^{(H_p \times W_p) \times C} \quad (8)$$

Therefore, as shown in Fig.3, the resultant weighted matrix is reshaped back and resized to obtain $X_{att,channel} \in \mathbb{R}^{H \times W \times C}$ feature tensor, which is given by:

$$X_{att,channel} = Unpooling(Reshape(X_{dr,att.ch}), p). \quad (9)$$

It should be noted that the prior Max-pooling layer ($p > 1$) in the SDA structure helps the overall attention process to be much faster and more space-efficient in practice than when $p = 1$.

*3) Self-adaptive Attention-based Skip Connection:* The SASC paths are adopted to replace the regular skip-connection (SC) between encoder and decoder counterpart blocks. Their process is based on the SDA module. The SASCs concatenate the decoder blocks' output feature maps with their SDA-processed versions, and the resultant tensors are passed to the decoder blocks for further processing. The SASCs provide a mechanism to adaptively highlight and transfer global information from encoder to decoder parts for better data reconstruction. Accordingly, lower α and larger β values for the SDAs' weighting parameters are adaptively assigned at each scale to capture richer contextual dependencies and model semantic interdependencies in the channel dimension. This guides the model to fuse multi-semantic global contextual features and improves the ability of the model to pay attention to global salient features and suppress the interference of irrelevant local features adaptively.

*4) Residual Multi-kernel Pooling:* Similar to [52], we propose to use the RMP block in the encoder part of the model which handles the large variety of fluid sizes in retinal OCT image representations. RMP involves multiple effective field-of-views for data processing and encodes global context information with different-sized receptive fields. In our model, the SAD and RMP modules are inserted into the top of the final encoder block to enhance latent-space representation potential.

*5) Decoder Part:* There are five decoder blocks in the decoder part, as illustrated in Fig.2. The key components of each decoder block include a Conv2DTranspose layer for up-sampling, a feature concatenation operation from the associated SASC path, and three consecutive blocks of Conv2D-BN-ReLU layers. The decoder part is primarily used to restore spatial information provided by SDA+RMP with strong multi-scale semantic features and to gradually fuse multi-semantic global contextual information from SASC paths.

*6) Deep Supervision Scheme and Loss Function*: In retinal OCT data analysis, the complicated pathological manifestations of retinal fluids with variations in size, shape, and location, and the presence of interferences such as other lesions, speckle noise, and also blurry boundaries for the suspected areas, pose great challenges to precisely segment fluid regions. Therefore, learning multi-scale non-local features along with key local features is essential for segmentation performance improvement. For this purpose, in addition to the previous model development considerations, we also proposed a DSL scheme for RetiFluidNet model training where misrepresentations on different scales are incorporated into the computation of the overall cost function. In this scheme, as shown in Fig.2, the FMs with different scale information from the middle decoder blocks are employed to generate auxiliary output tensors, which include standard convolutional outputs and Bicon output predictions at different scales. The tensors are then used with a scale-dependent loss weighting method to compute the DLC and CLC terms, while the weighting method regulates the impact of local losses for texture- and edge-specific penalties, respectively.

*(6-A) DLC term computation:* We used an up-sampling layer by a factor of P followed by a Conv2D layer with C filters, 1×1 kernel size, and a Softmax activation function after each decoder block. Here, P and C represent the scale-dependent up-sampling factor and the number of fluid classes, respectively. For each block, the size of the local FMs is compared to the size of the original ground truth mask, and P is chosen accordingly. Then, we integrated the decoder blocks' outputs with the original output dice loss into the DLC component in the proposed compound objective function for model training. First, we define the dice loss function:

$$L_{dice}(y_{pred}, y_{true}) = 1 - \frac{2\sum_{i,j}^{H,W}(y_{pred} \times y_{true})}{\sum_{i,j}^{H,W}(y_{pred}) + \sum_{i,j}^{H,W}(y_{true})}, \quad (10)$$

where H and W are the height and width of the input image, respectively. Thus, the DLC term, which is a weighted-sum version of the multi-scale dice loss components, is given by:

$$DLC = L_{dice}(y_{pred}, y_{true}) + \sum_{l=1}^{L} \frac{1}{2^{l-1}} L_{dice}(y_{pred,l}, y_{true}), \quad (11)$$

Here, $l$ indicates the index of auxiliary output, $y_{pred,l}$ is the $l^{th}$ auxiliary predicted output and $y_{true}$ is the annotated ground truth. Fig.2 illustrates more details about the local loss definitions. As seen from the figure, our proposed model mainly consists of $L = 4$ middle encoder blocks for auxiliary output generation, and $l_0$ indicates the original output scale.

*(6-B) CLC term computation:* To further incorporate edge information into our local loss computations, we employed a modified and multi-class version of the Bicon loss (BL) function proposed in [53]. Therefore, the CLC term is defined as a weighted-sum version of the multi-scale BL local loss components and can be expressed by:

$$CLC = BL(y_{pred}, y_{true}) + \sum_{l=1}^{L} \frac{1}{2^{l-1}} BL_{scale_l}(y_{pred,l}, y_{true}), \quad (12)$$

Here:

$$BL = \frac{1}{2} L_{decouple} + L_{con.map} + L_{con.dice}. \quad (13)$$

The first term in Eq.13, $L_{decouple}$, is the edge-decoupled loss and is computed using the categorical cross-entropy (CCE) loss between the edge-decoupled map $\tilde{S}_{decouple}$ and the ground truth segmentation mask $y_{true}$ in the one-hot coding format, given by:

$$L_{decouple} = L_{CCE}(\tilde{S}_{decouple}, Y_{true})$$
$$= \begin{cases} \sum_{c=1}^{C} L_{BCE}\left(1 - \min\{BCM_{k,c}(i,j)\}_{k=1}^{8}, y_{true,c}(i,j)\right); (i,j) \in P_{edge,c} \\ \sum_{c=1}^{C} L_{BCE}\left(\min\{BCM_{k,c}(i,j)\}_{k=1}^{8}, y_{true,c}(i,j)\right) \quad ; (i,j) \notin P_{edge,c} \end{cases} \quad (14)$$

In this equation, $L_{BCE}$ is the binary cross-entropy loss function, $BCM$ indicates the bilateral connectivity map [53], and $P_{edge,c}$ refers to the set of ground truth edge pixels that are obtained from the connectivity mask [56] for class $c$. In Eq.13, we use the following definition of connectivity-map loss $L_{con.map}$:

$$L_{con.map} = L_{CCE}(CM, Y_{true.con}) = \sum_{c=1}^{C} L_{BCE}(CM_c, y_{true.con,c}), \quad (15)$$

where, $CM$ and $y_{true.con}$ are the connectivity map and connectivity mask, respectively [53]. $y_{true.con}$ is generated from the original segmentation mask $y_{true}$ according to [53]. The third term in Eq.13, $L_{con.dice}$, is the connectivity-based dice loss, defied by Eq.10 when $y_{pred}$ is replaced by $\tilde{S}_{global}$. Here, $\tilde{S}_{global}$ is constructed by the maximum connection probability across main output channels [53].

Finally, we define the overall objective function of our network as:

$$Loss_{joint} = DLC + \lambda . CLC \quad (16)$$

Our proposed joint objective function in Eq.16 benefits from cross-scale and long-range dependencies among different blocks of encoders and decoders for loss computation. It is worth noting that, for a given input image, the connectivity-based segmentation output maps of the decoder block 1 are used to generate RetiFluidNet's inference output as in [53].

## IV. EXPERIMENTAL STUDY AND RESULTS

### A. Experimental Setup

In this section, we first introduce our strategy for data augmentation, and then implementation details are presented.

*1) Data Augmentation Strategy:* The image data was augmented with random translations, rotations, contrast adjustments, and mirroring, resulting in eight augmented images for each image in the training datasets.

*2) Implementation Details:* Our implementation is based on TensorFlow v2.4.0, and Python 3.7. All tests were run on a PC with NVidia GeForce Titan-V GPU (12GB), and 128 GB of RAM. For model training, the hyperparameters, including the maximum number of epochs (30), learning rate (2e-4), learning rate decay schedule (learning rate was multiplied by 0.8 every 5 epochs), batch size (4), and optimization method (the RMSprop optimizer) were the same for all of the baselines. In addition, the seed point for random data partitioning in the cross-validation method was preserved the same for all the networks to ensure the evaluation results have no bias due to the data splitting step. Throughout the experiments, the lambda hyperparameter was set to 0.05 and was optimized using the grid search method with a range of [0.01, 0.05, 0.1, 0.5, 1] over the RETOUCH training subsets. The code scripts will be available at: *https://github.com/aidialab/RetiFluidNet.*

### B. Performance Measures

The following evaluation measures were used to calculate the segmentation performance of different methods in this study. Dice similarity coefficient (DSC) and balanced accuracy (ACC) were used for evaluating the similarity between OCT images and ground truths and pixel-wise classification performance according to the following equations:

$$DSC = \frac{2|y_{pred} \cap y_{true}|}{|y_{pred}| + |y_{true}|}, \quad (17)$$

$$Acc = \frac{1}{2}\left(\frac{TP}{TP+FN} + \frac{TN}{TN+FP}\right), \quad (18)$$

For pixel classification, TP, TN, FP, and FN stand for true positive, true negative, false positive, and false negative, respectively. The segmentation result and ground truth are represented by $y_{pred}$ and $y_{true}$, respectively.

### C. Validation Method

To evaluate and generalize the performance of the deep learning models, we used the unbiased k-fold cross-validation (CV) method. Folding was performed at the subject level to ensure that the B-scans from the same OCT were not used in both the training and testing sets throughout each iteration. The model's final performance was obtained by averaging the evaluation metrics on testing sets over iterations.

### D. Study Design

The following studies were conducted in this work to demonstrate the competence of the proposed model and to obtain a benchmark for comparing the performance of RetiFluidNet in the retinal fluid segmentation task.

*1) Baseline Study:* We compared the segmentation performance of the proposed model to those of the common image segmentation CNN models, including U-Net [22], FCN [23], Seg-Net [24], Deeplabv3+ [25], APC-Net [54], DA-Net [48], DDR-Net [55], and also the recently developed medical image segmentation models including U-Net++ [26], Attention U-Net [33], CE-Net [52], CPF-Net [35], nnU-Net [56], and MsTGANet [36]. All the networks were trained using the same configuration to ensure fairness in the evaluation step. The cross-validated results on the RETOUCH challenge dataset are reported in Table 3.

*2) State-of-the-art Analysis:* The proposed method was also compared to several recently developed fluid region segmentation algorithms, including patch-based U-Net segmentation method [3]; multi-scale and dual attention enhanced nested U-Net (MDAN-UNet) segmentation model [27]; multi-label deep supervision segmentation method [42]; attention-based U-Net segmentation approach [34]; generative adversarial network (GAN)-based segmentation-renormalized image translation framework [43], and retinal fluids segmentation network (RFS-Net) proposed in [47]. For fair comparisons, the target datasets, pre/post-processing pipelines, and evaluation methodologies of the original studies were meticulously followed. The comparative results are summarized in Table 4.

*3) Ablation Study:* In the proposed method, the SDA attention module, the SASC paths, and the multi-scale DSL scheme are embedded into the basis U-Net architecture. To further understand the influence of each component in our model, we conducted a comprehensive ablation study, and the experimental results are reported in Table 5. The customized U-Net (cU-Net) model used in RetiFluidNet was adopted as our baseline model to evaluate the effectiveness of the SDA, SASC, and DSL scheme. The cU-Net is the backbone of the proposed RetiFluidNet model in Fig.2, whereas the SDA modules and DSL scheme are eliminated from the RetiFluidNet structure, and the SASC paths are replaced by simple SCs similar to the standard U-Net [33]. In the first step, we explored the contribution of the attention module to the cU-Net performance by replacing the proposed SDA (in Fig.2) with the competitive squeeze and excitation (SqEx) [49], squeeze-and-attention (SqAt) [50], adaptive context module (ACM) [54], or dual-attention (DA) [48] mechanisms. The attention-based SCs were also modified accordingly. For this experiment, the loss function in Eq. 10 was considered for the optimization process. Next, the DSL scheme contribution was analyzed on the best contributing model. For this purpose, different combinations of local DLC, CLC, or joint losses were evaluated. The ablation experiments were carried out on the RETOUCH dataset using the 3-fold CV method.

*4) Statistical Analysis:* The statistical significance of the performance improvement for the proposed RetiFluidNet versus the best alternative models was evaluated using the Wilcoxon signed-rank test. We determined the *p*-values for comparisons of the DSC metric, in which ten repetitions of the 3-fold CV method were employed for statistical significance assessment. Table 6 shows the results of the statistical analysis.

## V. DISCUSSION

### A. Quantitative Evaluation

The mean and standard deviation values of the DSC and balanced ACC measures were used to quantitatively evaluate the retinal fluid segmentation performance on the RETOUCH dataset in the baseline study. As shown in Table 3, our proposed model outperformed the comparative baseline methods, with average DSC and ACC values of 86.6% and 94.9%, respectively, over different fluid types and scanners. When compared to MsTGANet, the best performer of all baseline methods, the average DSC and ACC values for RetiFluidNet were 3.1% and 0.7% higher, respectively. Despite the fact that the transformer-based model of MsTGANet still learns features by capturing long-distance feature correlation across the entire FMs, the RetiFluidNet model's sufficient usage of inter-pixel information results in better segmentation performance in near-edge regions and low spatial coherence. Due to the embedding of the DSL scheme, RetiFluidNet took slightly longer to run than MsTGANet. It is, however, sufficient to meet the requirement for near-real-time image inferencing.

Several different retinal fluid segmentation methods were investigated to provide a more comprehensive assessment of the performance of the proposed method. The experimental results in Table 4 demonstrate that, when compared to existing state-of-the-art approaches for retinal OCT fluid segmentation, the proposed RetiFluidNet achieves superior segmentation performance due to the proposed attention modules, the DSL scheme, and the structure of the overall model.

For the sake of completeness, we also performed a comprehensive ablation study on the proposed structure of the RetiFluidNet; the results are shown in Table 5 (*Attention Module Analysis* section). When cU-Net+SDA+SA.SC+DSC is compared to cU-Net+DSC, the average DSC improves from 73.5% to 85.3%, as the proposed SDA module and SASC paths can adaptively guide the model to learn multiscale local/non-local features with long dependency information. Additionally, we conducted experiments to compare the performance of the SqEx [49], SqAt [50], ACM [54]*,* and DA [48] attention modules. When compared to cU-Net+SqEx+SqEx.SC, cU-Net+SqAt+SqAt.SC, cU-Net+ACM+ACM.SC, and cU-Net+DA+DA.SC our proposed cU-Net+SDA+SA.SC achieves superior performance in DSC indices (with 9.1%, 7.9%, 2.6%, and 3.1% increases, respectively) while maintaining the same time efficiency, demonstrating the effectiveness of the proposed SDA module with adaptive fusion. Furthermore, Table 5 (*DSL Scheme Analysis* section) demonstrates that incorporating the DSL scheme into the cU-Net improved segmentation performance, with cU-Net+Multi-scale-DLC outperforming cU-Net in terms of the DSC indicator. When the CLC term was further integrated into the objective function, the DSC of cU-Net+Multi-scale-JointLoss was improved from 73.5% to 80.5%. These results further indicate that the proposed

TABLE 3. BASELINE PERFORMANCE EVALUATION OVER RETOUCH DATASET USING DSC & ACC METRICS AND THE 3-FOLD CROSS-VALIDATION METHOD

| Scanner | Metrics | Fluid Type | Baseline Networks ||||||||||||| 
|---|---|---|---|---|---|---|---|---|---|---|---|---|---|---|---|
| | | | U-Net | FCN | Seg-Net | Deeplabv3+ | APC-Net | DA-Net | DDR-Net | U-Net++ | Attention U-Net | CE-Net | CPF-Net | nnU-Net | MsTGANet | RetiFluidNet |
| Cirrus | DSC | IRF | 52.0±19.9 | 69.8±7.5 | 68.6±7.1 | 83.3±9.9 | 69.8±7.6 | 67.9±6.7 | 76.4±4.5 | 73.7±5.4 | 76.9±5.1 | 77.3±3.4 | 79.9±3.1 | 78.7±2.5 | 80.0±**2.1** | **84.5**±2.6 |
| | | SRF | 73.5±20.7 | 77.3±8.8 | 76.6±8.4 | 88.9±6.6 | 77.4±8.8 | 76.4±7.8 | 79.3±7.6 | 81.6±5.6 | 81.1±6.3 | 80.9±4.2 | 88.6±2.2 | 89.4±**0.5** | 88.3±2.7 | **90.7**±2.2 |
| | | PED | 72.7±21.7 | 79.8±5.9 | 78.5±5.4 | 84.5±10.5 | 79.8±5.9 | 79.2±6.1 | 82.5±10.0 | 79.2±7.8 | 74.3±11.1 | 74.9±12.0 | 82.6±**4.7** | 80.9±8.4 | 83.1±8.5 | **85.4**±5.9 |
| | ACC | IRF | 85.6±4.2 | 84.7±3.8 | 84.7±3.8 | 93.0±3.2 | 84.7±3.8 | 81.4±3.9 | 86.1±2.6 | 89.9±2.8 | 92.8±3.2 | **94.3**±**0.9** | 92.9±1.2 | 93.1±1.2 | 93.1±1.7 | 93.5±2.3 |
| | | SRF | 90.4±5.3 | 88.6±4.4 | 88.6±4.4 | 96.4±1.3 | 88.6±4.4 | 87.5±6.4 | 88.2±4.6 | 91.6±1.2 | 95.9±3.6 | 95.9±1.8 | 95.5±2.1 | 96.2±**0.8** | 95.1±1.3 | **96.7**±1.2 |
| | | PED | 90.4±5.3 | 89.8±2.9 | 89.8±2.9 | **95.1**±2.2 | 89.8±2.9 | 90.8±3.6 | 90.1±5.4 | 90.1±1.2 | 93.1±2.4 | 93.3±2.9 | 93.1±2.4 | 93.5±**0.2** | 94.4±2.1 | 94.2±2.0 |
| Spectralis | DSC | IRF | 61.4±12.3 | 57.5±3.1 | 44.3±12.1 | 74.4±2.2 | 57.5±3.1 | 63.0±9.4 | 70.4±**1.4** | 70.9±5.9 | 68.4±6.8 | 66.6±14.9 | 75.2±7.7 | 75.5±3.6 | 78.7±1.8 | **82.4**±2.9 |
| | | SRF | 75.3±12.1 | 71.2±5.9 | 62.8±16.9 | 83.6±3.6 | 71.2±5.9 | 75.4±11.3 | 84.0±6.3 | 84.3±1.7 | 79.9±8.3 | 79.6±7.3 | 86.1±4.0 | 88.7±2.5 | 88.4±2.1 | **93.2**±**1.5** |
| | | PED | 80.5±8.5 | 74.5±4.1 | 62.0±15.0 | 79.3±6.9 | 74.5±4.1 | 76.4±5.1 | 75.8±8.6 | 80.2±3.2 | 76.5±5.3 | 77.5±**3.0** | 81.8±6.4 | 81.6±4.6 | 84.5±4.5 | **87.6**±3.6 |
| | ACC | IRF | 81.4±7.3 | 78.6±1.5 | 79.6±2.6 | 90.5±1.9 | 78.6±1.6 | 82.8±6.6 | 82.2±**0.5** | 91.3±1.1 | 90.6±2.8 | 90.4±2.4 | 91.5±1.4 | 91.7±0.8 | **92.2**±1.1 | **92.2**±1.0 |
| | | SRF | 87.8±6.4 | 85.5±2.9 | 85.7±2.8 | 96.8±0.7 | 85.6±3.0 | 88.5±6.8 | 90.1±3.5 | 96.8±1.0 | 97.2±2.9 | 97.0±1.0 | 97.2±**0.6** | 96.9±**0.6** | **97.3**±0.9 | **97.3**±0.9 |
| | | PED | 90.7±4.9 | 87.1±2.5 | 87.5±2.8 | 93.3±0.3 | 87.2±2.6 | 90.2±3.6 | 86.1±4.6 | 94.3±0.8 | 94.0±1.7 | 94.3±0.8 | 93.6±0.9 | 94.4±1.2 | 94.8±1.0 | **95.6**±**0.5** |
| Topcon | DSC | IRF | 74.2±3.1 | 74.4±5.4 | 73.0±5.7 | 79.6±4.7 | 79.9±3.5 | 79.7±**2.3** | 80.7±5.6 | 77.7±7.3 | 76.9±3.2 | 76.2±3.6 | 77.8±6.9 | 78.4±7.0 | 78.7±5.5 | **80.9**±5.7 |
| | | SRF | 85.1±7.3 | 85.3±10.6 | 77.7±**4.3** | 84.7±6.8 | 83.7±7.1 | 85.3±11.4 | 85.6±6.9 | 83.3±6.7 | 85.4±6.1 | 81.5±8.0 | 82.7±8.5 | 83.8±7.9 | 85.7±7.2 | **87.9**±6.8 |
| | | PED | 80.2±8.9 | 80.6±10.4 | 80.6±10.4 | 76.6±5.8 | 86.2±5.8 | 81.9±8.8 | 81.4±5.7 | 82.0±10.4 | 79.1±5.9 | 77.7±7.9 | 82.8±**2.1** | 82.1±6.0 | 84.4±4.7 | **87.1**±5.3 |
| | ACC | IRF | 87.0±1.6 | 87.0±2.7 | 87.1±2.7 | 93.0±1.6 | 93.9±**0.8** | 89.2±1.2 | 88.8±3.2 | 93.4±1.9 | 90.3±1.9 | 89.3±0.8 | 93.4±3.7 | 93.9±1.1 | 92.9±1.7 | **94.6**±1.1 |
| | | SRF | 92.5±3.7 | 92.6±5.3 | 92.5±5.3 | 95.2±4.2 | 95.4±4.6 | 92.6±5.7 | 92.1±**1.0** | 95.2±4.6 | 94.9±4.0 | 93.8±5.8 | 94.8±5.0 | 95.5±4.4 | 94.7±4.2 | **95.7**±4.2 |
| | | PED | 90.4±4.4 | 90.2±5.2 | 90.2±5.2 | 93.6±3.8 | **94.4**±3.2 | 92.3±4.4 | 89.4±**1.3** | 92.9±4.4 | 92.7±4.1 | 92.2±3.5 | 93.8±3.2 | 94.1±3.3 | 93.7±2.7 | 94.3±2.8 |
| Test time (sec/B-scan) || | 0.39 | 0.41 | 0.34 | 0.23 | 0.39 | 0.28 | 0.12 | 0.33 | 0.40 | 0.20 | 0.13 | 0.18 | 0.31 | 0.44 |

TABLE 4. STATE-OF-THE-ART PERFORMANCE EVALUATION OVER STUDY DATASETS

| References | Method Configuration ||| Validation Method | Dataset | Metric | Ave-Results* (%) | Our Ave-Results* (%) |
|---|---|---|---|---|---|---|---|---|
| | Input Size | Pre-Process | Post-Process | | | | | |
| Alsaih et al. 2020 [3] (DeeplabV3+$_{Patch}^{Ex}$) | 384×384 | BM3D Denoising, Image Patching (128×128) | Median Filtering, Image Closing Operation | 3-F CV | RETOUCH OPTIMA DUKE | DSC | IRF,SRF,PED=84.3,82.3,81.3 IRC=70.5 IRC=80.0 | IRF,SRF,PED=86.6,91.8,88.3 IRC=82.6 IRC=83.5 |
| W.Liu et al. 2020 [27] (MDAN-UNet-32) | 512×400 | Normalization, Median Filtering | - | H.O | RETOUCH | DSC | IRF,SRF,PED=69.3,66.8,64.8 | IRF,SRF,PED=76.6,83.7,79.8 |
| X.Liu et al. 2021 [34] | 256×256 | CNN-based Layer Segmentation | Mask Thresholding | 4-F CV | OPTIMA | DSC | IRC=82.8 | IRC=83.9 |
| Reis et al. 2021 [42] ** | 200×200 | Anisotropic Filtering | - | 10-F CV | RETOUCH | Mean IoU | Cirrus:51.2, Topcon:42.7 | Cirrus:53.7, Topcon:46.6 |
| Ren et al. 2021 [43] ** | 256×256 | - | - | H.O | RETOUCH | DSC | IRF,SRF,PED=67.1,62.3,52.5 | IRF,SRF,PED=68.9,65.1,54.3 |
| Hassan et al. 2021 [47] | 360×480 | NLM Filtering, Total Retina Segmentation | - | H.O | RETOUCH | DSC | IRF,SRF,PED=76.1,79.6,80.4 | IRF,SRF,PED=83.0,86.9,85.2 |

\* The reported results were calculated using the average DSC metric across all scanners' datasets. The pre/post-processing and validation methods and the target datasets used in the reference papers were considered for the analyses.
\*\* For further information on the experimental setup considered for the comparison, please refer to the main paper. NLM=Non-local Mean.

TABLE 5. ABLATION STUDY RESULTS OVER RETOUCH DATASET USING DSC METRIC AND THE 3-FOLD CROSS-VALIDATION METHOD

| | Method | Results* (%) ||||
|---|---|---|---|---|---|
| | | IRF | SRF | PED | Ave. |
| Baseline | cU-Net + DSC Loss | 64.3±12.9 | 78.7±13.7 | 77.5±12.8 | 73.5±13.2 |
| Attention Module Analysis | cU-Net + SqEx [49] + SqEx.SC + DSC Loss | 69.9±8.8 | 79.1±10.1 | 79.6±8.3 | 76.2±9.1 |
| | cU-Net + SqAt [50] + SqAt.SC + DSC Loss | 68.4±8.1 | 83.5±8.9 | 80.3±6.9 | 77.4±8.0 |
| | cU-Net + ACM [54] + ACM.SC + DSC Loss | 80.4±8.5 | 85.1±8.7 | 82.5±5.8 | 82.7±7.7 |
| | cU-Net + DA [48] + DA.SC + DSC Loss | 78.2±6.2 | 86.7±5.8 | 81.6±5.3 | 82.2±5.8 |
| | cU-Net + SDA + SC + DSC Loss | 78.1±6.7 | 89.0±6.8 | 80.8±7.6 | 82.6±7.0 |
| | cU-Net + SDA + SA.SC + DSC Loss | 81.8±6.1 | 90.2±6.5 | 83.9±5.7 | 85.3±6.1 |
| DSL Scheme Analysis | cU-Net + Multi-scale DLC | 69.4±9.8 | 79.4±9.0 | 78.7±8.4 | 75.9±9.1 |
| | cU-Net + Multi-scale Joint Loss | 75.3±5.9 | 84.9±6.2 | 81.2±6.1 | 80.5±6.1 |
| | cU-Net + SDA + SA.SC + Multi-scale DLC | 85.0±5.7 | 92.7±6.0 | 87.3±6.4 | 88.4±6.1 |
| | cU-Net + SDA + SA.SC + Multi-scale CLC | 87.7±2.9 | 91.3±**2.8** | 89.5±**2.1** | 89.5±**2.7** |
| | cU-Net + SDA + SA.SC + Multi-scale Joint Loss: *RetiFluidNet* | **91.1**±2.8 | **93.6**±3.7 | **92.8**±2.9 | **92.5**±3.2 |

\* The reported results were calculated by combining all the RETOUCH OCTs acquired from different scanners for model training and evaluation. X.SC=Attention-based Skip Connection exploiting the X module.

TABLE 6. STATISTICAL ANALYSIS OF THE PERFORMANCE OF THE PROPOSED RETIFLUIDNET VERSUS ALTERNATIVE METHODS

| | Compared Methods | P-value ||||
|---|---|---|---|---|---|
| | | IRF | SRF | PED | IRC* |
| RetiFluidNet vs. Baseline Methods | Deeplabv3+ [25] | 0.026 | <0.001 | <0.001 | - |
| | U-Net++ [26], DA-Net [48], CE-Net [52], APC-Net [54], DDR-Net [55] | <0.001 | <0.001 | <0.001 | - |
| | CPF-Net [35] | <0.001 | <0.001 | 0.015 | - |
| | MsTGANet [36] | 0.033 | 0.027 | 0.041 | - |
| | nnU-Net [56] | <0.001 | 0.019 | 0.003 | - |
| RetiFluidNet vs. Fluid Seg. Methods | Alsaih et al. [3] | 0.088 | <0.001 | 0.002 | <0.001 |
| | X.Liu et al. [34] | - | - | - | 0.038 |
| | Reis et al. [42] | <0.001 | <0.001 | <0.001 | - |
| | Ren et al. [43] | 0.057 | <0.001 | <0.001 | - |
| | Hassan et al. [47] | <0.001 | <0.001 | <0.001 | - |
| | RetiFluidNet without CLC | 0.051 | 0.021 | 0.003 | - |
| | RetiFluidNet without DLC | 0.033 | 0.040 | 0.069 | - |

\* The average DSC metric for the mixed OPTIMA and DUKE datasets was used to estimate the *p*-values.

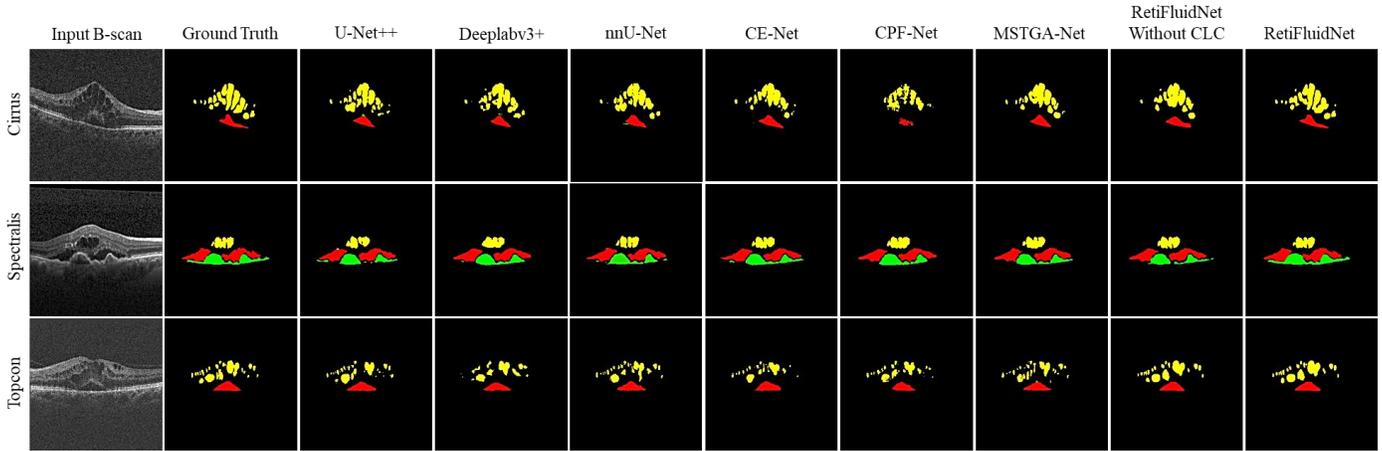

Fig 4. Comparison of different method segmentation results: visualization using baseline models and RetiFluidNet over RETOUCH test datasets. The IRF, SRF, and PED fluid segmentations are shown in yellow, red, and green, respectively.

DSL scheme considerably improves the performance of the model. Additionally, as shown in Table 5, the segmentation performance of the RetiFluidNet is significantly better than that of cU-Net+Multi-scale-JointLoss: the average DSC for our model improved from 75.3%, 84.9%, and 81.2% to 91.1%, 93.6%, and 92.8% respectively, for IRF, SRF, and PED segmentation. Compared with RetiFluidNet without CLC, the DSC of RetiFluidNet improved considerably by 6.1%, 0.9%, and 5.5% for IRF, SRF, and PED segmentation, respectively. RetiFluidNet also outperformed the version that only used CLC (without the DLC), with the DSC of RetiFluidNet increasing on average by 3% ($p$-value$<0.05$) for fluid segmentation. These results demonstrate that the proposed RetiFluidNet with multi-scale joint-loss can significantly improve the fluid segmentation performance by further leveraging the edge-connectivity-based information and the key textural and contextual information.

### B. Qualitative Assessment

Fig.4 shows sample segmentation results for the proposed RetiFluidNet on the RETOUCH test datasets and comparisons with six high-performance baseline networks and the RetiFluidNet without CLC. The IRF, SRF, and PED fluid segmentations are depicted in yellow, red, and green, respectively. As shown in Fig.4, the proposed model obtains higher segmentation performance, especially when segmenting small-sized retinal fluids. The competitive U-Net++ model (top and middle rows) suffers from false positives caused by the interference of non-correlated local information and speckle noise. Moreover, the proposed RetiFluidNet achieves better segmentation performance than nnU-Net, CE-Net, MsTGANet, and RetiFluidNet without the CLC term, in which speckle noise interferences and blurry edges negatively affect the results. Taken together, these results demonstrate the effectiveness and robustness of our proposed method.

### C. Statistical Comparison

The statistical significance of the performance improvement for the proposed RetiFluidNet versus the best alternative models is shown in Table 6. All gains in DSC for RetiFluidNet (compared to other methods) are statistically significant with $p$-values less than 0.05, except for the methods described by Alsaih et al. [3], and Ren et al. [43] for IRF segmentation. These findings support the effectiveness of the proposed attention module that was embedded in the RetiFluidNet. The table also reports the $p$-values of the suggested DSL scheme in comparison to the other competitive alternatives in the ablation study. All DSC improvements for RetiFluidNet are statistically significant, with $p$-values less than 0.05, with the exception of RetiFluidNet without CLC for IRF segmentation and RetiFluidNet without DLC for PED analysis. These results demonstrate that the suggested DSL scheme would significantly improve segmentation performance by leveraging edge-connectivity information for multi-scale representing fluid regions. Comparison to other segmentation models showed that the proposed RetiFluidNet benefits from hierarchical representation learning of textural, contextual, and edge features using the SDA module, multiple SASC connections, and the multi-scaled DSL scheme. The attention mechanism of the proposed SDA module enables the model to automatically extract deformation-aware representations at multiple levels, and the introduced SASC paths take spatial-channel interdependencies into account when concatenating counterpart encoder and decoder units, thereby improving local and global representational capability. The SDA's contribution is to highlight the spatial or channel information adaptively for better representation. The $\alpha$ and $\beta$ values are optimized at each processing block/scale to adaptively capture cross-pixel and cross-channel dependencies in order to effectively highlight local and global structures for more precise segmentation performance.

Additionally, RetiFluidNet was optimized using a joint loss function that incorporated a weighted version of dice overlap and edge-preserved connectivity-based losses as well as multiple hierarchical levels of multi-scaled local losses, all of which resulted in improved segmentation.

### D. Combine with Pre-trained CNN Backbones

RetiFluidNet can be combined with any convolutional segmentation backbone. As an alternative backbone, we also employed ResNet50 for the encoder block of the model, which is frequently used in natural image semantic segmentation. The experimental results showed that the segmentation performance of res50-RetiFluidNet in the best training mode (i.e., fine-tuning) is not significantly superior to that of RetiFluidNet on the RETOUCH dataset. RetiFluidNet and res50-RetiFluidNet have shown comparable average dice performance of 92.5%±3.2 and 92.2%±2.1, respectively. The fact that the RETOUCH dataset was acquired from three different imaging devices is a challenging factor. It is difficult for the ResNet50 encoder block to manage fluid images from different data sources with varying shapes, while the OCT fluid segmentation datasets are quite small compared to other natural image segmentation datasets. Therefore, utilizing a pre-trained CNN model with a large number of parameters as the encoder on a limited-size dataset may result in performance degradation due to overfitting. Therefore, to apply our methodology to a real-world dataset, one can therefore select a suitable backbone network to adapt to various datasets. In future work, we plan to develop a 3-D version of our model for volumetric fluid region segmentation and accelerate the multi-class training phase of our proposed segmentation network.

## VI. Conclusion

In this study, we proposed a deep learning framework called the RetiFluidNet model for segmenting different types of fluid in macular OCT images. We trained and evaluated this model on publicly available datasets acquired on different OCT scanners. RetiFluidNet is well-suited for clinical applications where quantification of retinal fluids is required by providing a fast and robust automated technique for precise fluid segmentation. Experimental results showed that RetiFluidNet requires minimal image pre-processing steps, exhibits high performance in terms of Dice and ACC metrics, and is a reliable method for segmenting retinal OCT images even in the presence of disruptive speckle noise and severe pathologies.